\documentclass[aip,pre,preprint,superscriptaddress,showpacs,floatfix]{revtex4-1}

\usepackage{blindtext}
\usepackage{float}
 \usepackage{graphicx}
 \usepackage{siunitx}
 \usepackage{xcolor}
\usepackage{amssymb}
\usepackage{amsmath}

\begin{document}
\title{Phenomenon of self-oscillation in bubble dynamics: Bouncing acoustic bubbles}
\author{Gabriel Regnault}
\affiliation{Univ Lyon, École Centrale de Lyon, INSA de Lyon, CNRS, LMFA UMR 5509, F-69002 Lyon, France}
\author{Alexander A. Doinikov}
\affiliation{Univ Lyon, École Centrale de Lyon, INSA de Lyon, CNRS, LMFA UMR 5509, F-69002 Lyon, France}
\author{Gabrielle Laloy-Borgna}
\affiliation{Univ Lyon, Université Claude Bernard Lyon 1, Centre Léon Bérard, INSERM, UMR 1032, LabTAU, F-69003 Lyon, France}
\author{Cyril Mauger}
\affiliation{Univ Lyon, École Centrale de Lyon, INSA de Lyon, CNRS, LMFA UMR 5509, F-69002 Lyon, France}
\author{Philippe Blanc-Benon}
\affiliation{Univ Lyon, École Centrale de Lyon, INSA de Lyon, CNRS, LMFA UMR 5509, F-69002 Lyon, France}
\author{Stefan Catheline}
\affiliation{Univ Lyon, Université Claude Bernard Lyon 1, Centre Léon Bérard, INSERM, UMR 1032, LabTAU, F-69003 Lyon, France}
\author{Claude Inserra}
\email{claude.inserra@inserm.fr}
\affiliation{Univ Lyon, Université Claude Bernard Lyon 1, Centre Léon Bérard, INSERM, UMR 1032, LabTAU, F-69003 Lyon, France}

\begin{abstract}
Self-oscillations underlie many natural phenomena such as heartbeat, ocean waves, and the pulsation of variable stars. From pendulum clocks to the behavior of animal groups, self-oscillation is one of the keys to the understanding of synchronization phenomena and hence the collective behavior of interacting systems. In this study, we consider two closely spaced bubbles pulsating in the kHz range in response to ultrasonic excitation. A translational bouncing motion emerges from their interaction with a much lower frequency than the bubble pulsation frequency. Our analysis reveals that the observed bubble bouncing exhibits the main features of self-oscillation, such as negative damping and the emergence of a limit cycle. These results highlight unexpected nonlinear effects in the field of microbubbles and give insights into the understanding of synchronization in large bubble clouds.
\end{abstract}

\maketitle

\textit{Introduction}~--- 
Unlike forced and parametric oscillations, self-oscillation is a periodic motion that is generated and maintained without an external force that explicitly depends on time with a similar or multiple periodicity~\citep{Andronov,Landa}. Forced and parametric resonances are very familiar to physicists, whereas self-oscillation effects are little known despite their widespread occurrence in nature and engineering. Self-oscillatory systems are found in mechanics, electronics, acoustics, biomechanics, and even in finance and macroeconomics. Heartbeat, clocks, bowed and wind instruments (including the human voice), explosion engines, ocean waves, fluttering of tree leaves, and the pulsation of variable stars (cepheids)~\citep{Jenkins,Landa_2,Zeng} are examples of self-oscillatory phenomena. It is interesting to mention that the wind-powered “galloping” (torsional oscillation) of suspension bridges, as was in the famous case of the first suspension bridge over the Tacoma Narrows (part of Puget Sound in the U.S. state of Washington), which caused its collapse in 1940~\citep{Jenkins}, often wrongly attributed to forced resonance, is in fact self-oscillation. The variety of self-oscillatory systems make their study important from both applied and fundamental aspects. Furthermore, the ability of self-oscillatory systems to synchronize, as was first shown by Huygens for a pair of pendulum clocks~\citep{Pikovsky}, can be used to reveal important characteristics of complex systems and help, for instance, in the understanding of the collective behavior of animals~\citep{Vicsek}. Gas bubbles are remarkable objects that appear in many life and natural sciences. They contribute to effects such as the sound of rain~\citep{Prosperetti}, the foam production on sea waves~\citep{Veron}, and the tree embolism~\citep{Vincent}. When periodically driven by an acoustic wave, the so-called acoustic bubbles are the key elements in many engineering applications such as surface cleaning~\citep{Reuter}, ultrasound-mediated drug delivery~\citep{Lentacker} and the design of microrobots~\citep{Bertin}. The dynamics of an acoustic bubble exhibits many nonlinear features~\citep{Lauterborn} such as nonlinear resonances, chaotic oscillations and hysteretic behavior, typical of many nonlinear dynamical systems. When one acoustic bubble oscillates in the presence of another bubble, they interact through a radiation force~\citep{Bjerknes}, called the secondary Bjerknes force. Depending on the bubble pair properties, this force can be attractive or repulsive: If the oscillations are in phase, the bubbles attract each other; if not, they repel each other~\citep{Ida}. However, in the case of bubbles driven above their resonance frequencies and exposed to a high-amplitude sound field, the sign reversal of the radiation interaction force between the bubbles has been theoretically predicted~\citep{Zabolotskaya,Doinikov95,Pelekasis}, which can lead to a small separation distance at which the two bubbles can be maintained stable. Stable bubble pairs have been observed in bubble systems from the micrometric~\citep{Miller,Rabaud} to the millimetric~\citep{Barbat} range, but always at large interbubble distances. Recently, Regnault et al.~\citep{Regnault_PoF} have controlled the approach of two micrometric bubbles in a two-frequency levitation chamber and observed the stabilization of a bubble pair at an interbubble distance smaller than the mean bubble radii. This effect, which was attributed to the sign reversal of the radiation interaction force, was obtained above a threshold pressure amplitude. The behavior of the bubble pair for pressure amplitudes above this threshold was not investigated so far. 
Here, we report for the first time a manifestation of self-oscillation in bubble dynamics. When investigating the motion of a pair of acoustically excited gas bubbles ($\sim$~\SI{150}{\micro\meter} in radius at rest), we observed that two closely spaced bubbles (minimum distance between the bubble centers $\sim$~\SI{315}{\micro\meter}) pulsating in the kHz range (\SI{31.9}{\kilo\hertz}) experienced a translational motion occurring in the form of periodic convergence and rebound with a much lower frequency ($\sim$~\SI{400}{\hertz}). \textcolor{red}{}. Our analysis reveals that the observed bubble bouncing exhibits the main features of self-oscillation, including negative damping and the emergence of limit cycles. This is the first time that self-oscillation is observed in bubble dynamics, which reveals that the universality and the prevalence of this phenomenon are wider than previously thought.

\begin{figure}[t!]
\includegraphics[width=\columnwidth]{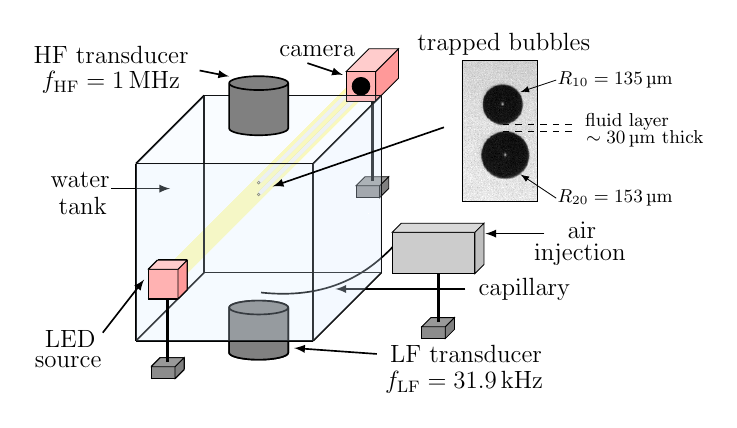}
\caption{\label{fig:fig1}{Schematic of the experimental setup. Bubbles are injected in a water tank by a capillary supplied by an air compressor, trapped at successive pressure nodes of a high-frequency (HF) standing acoustic wave, and then set into pulsations by a low-frequency (LF) ultrasound transducer. When the applied LF pressure exceeds a certain value, the bubbles jump to a new equilibrium position where they are almost in contact. The bubble dynamics is tracked with a high frame rate camera. The insert shows a snapshot of two bubbleswith equilibrium radii of $R_{10}=$\SI{153}{\micro\meter} and  $R_{20}=$\SI{135}{\micro\meter}, separated by a fluid layer of thickness of $\sim$~\SI{30}{\micro\meter}.}}
\end{figure}

\textit{Experimental setup}~--- 
A detailed description of our experimental setup (see Fig.~\ref{fig:fig1}) can be found in our previous works~\citep{Regnault20}. Microfiltered and demineralized water (Carlo Erba, water for analysis) was contained in a cubic, $8$-cm-edge, tank. Bubbles were generated at the tip of a capillary (inner diameter \SI{25}{\micro\meter}), immersed in the tank, linked to a microfluidic pressure controller (Elveflow, OB1 MK3) and supplied by an air compressor (Newport, ACWS). An uprising, periodically-spaced train of bubbles (mean radius \SI{100}{\micro\meter}, inter-bubble distance \SI{1}{\milli\meter}) was therefore established. Bubbles were trapped at successive pressure nodes of a standing acoustic wave induced by a high-frequency (HF) ultrasound transducer (Sofranel, IDMF018, $f_{\textrm{HF}}=$~\SI{1}{\mega\hertz}). Since the bubble equilibrium radii ($130–160$~\SI{}{\micro\meter}) were larger than the resonant radius at \SI{1}{\mega\hertz} ($R_{\textrm{res}}^{\textrm{HF}}=$\SI{3.73}{\micro\meter}), the primary Bjerknes force caused by the HF field~\cite{Regnault20} made the bubbles settle at neighboring pressure nodes located \SI{750}{\micro\meter} apart from each other. In view of the large difference between the HF resonant radius and the bubbles’ equilibrium radii, they did not experience any significant radial oscillations and were considered to be trapped at rest. The investigation of a unique bubble pair required to select only two successive trapped bubbles, by removing the others by hand. The bubble oscillations are induced by a low-frequency (LF) transducer (Sinaptec, $f_{\textrm{LF}}=$\SI{31}{\kilo\hertz}) located at the bottom of the tank and driven by a continuous sinusoidal signal. Due to the large difference between the LF wavelength ($\lambda_{\textrm{LF}}=$\SI{5}{\centi\meter}) and the inter-bubble distance, the bubble pair was assumed to be driven by a uniform pressure field. The dynamics of the bubble pair was captured with a CMOS camera (Vision Research V12.1) equipped with a $12\times$ objective lens (Navitar). Imaging was performed at $130 \times 10^3$ fps (frame size $128 \times 256$ pixels, 1 pixel $\sim$~\SI{5}{\micro\meter}) with backlight illumination. 

\textit{Observation of bouncing bubbles}~--- 
A controlled convergence of two bubbles towards each other is achieved by a step-by-step increase in the driving low-frequency (LF) acoustic pressure (see Fig.~\ref{fig:fig2}a). This convergence is induced by the interaction force between two bubbles, also called the secondary (Bjerknes) radiation force. Above a certain value of the applied pressure amplitude, the bubbles jump to a new equilibrium location, where they are separated by a thin fluid layer whose thickness is much smaller than the bubble radii. This position was observed only if both bubbles were larger than the resonance radius at \SI{31.9}{\kilo\hertz} (for this frequency the resonance radius of a single bubble is $R_{\textrm{res}}^{\textrm{LF}}=$\SI{103.4}{\micro\meter}). This quasi-contact location of the bubble pair results from the sign reversal of the radiation interaction force~\cite{Regnault_PoF} caused by multiple rescattering effects. A further increase in the LF pressure amplitude triggered the bouncing motion of the bubbles with a much lower frequency than the frequency of the LF driving field (Fig.~\ref{fig:fig2}b (Multimedia available online)). For a bubble pair with equilibrium radii of $R_{10}=$\SI{153}{\micro\meter} and $R_{20}=$\SI{135}{\micro\meter}, the bouncing frequency $f_{\textrm{bounce}}$ is \SI{403}{\hertz}. Note that for all the following results, the index “1” refers to the bottom bubble and the index “2” to the top bubble. The bouncing behavior remains remarkably stable throughout the observation time ($\sim1$ min, corresponding to $25000$ bouncing cycles). The distance between the centers of the bouncing bubbles $d(t)=z_2 (t)-z_1 (t)$, where $z_j (t)$ is the position of the center of the $j$th bubble along the line joining the bubble centers (Figs.~\ref{fig:fig2}a and ~\ref{fig:fig2}c), ranges from \SI{315}{\micro\meter} to \SI{405}{\micro\meter}, so the separation at the moment of the bubble convergence is as small as $1.09 (R_{10}+R_{20})$. Two approaching bubbles can coalesce or not depending on their Weber number based on their approach velocity~\citep{Duineveld1998}. This Weber number is defined as $\textrm{We}=\rho V^2 R_{eq} / \sigma$, where $\rho$ is the liquid density, $V$ is the velocity of approach of the two bubbles, $R_{eq}$ is the equivalent radius of the bubble pair, and $\sigma$ is the surface tension. An inhibition of the coalescence process has been observed when the Weber number exceeds the critical value $\textrm{We}=0.18$, a value that remains identical in the case of bubbles exposed to an acoustic field~\citep{Duineveld1996}. During the bouncing motion, for the mean equilibrium radius $R_{eq}=$\SI{150}{\micro\meter} and the approach velocity $V\sim$\SI{0.3}{\meter\second^{-1}}, the experimental Weber number in the bouncing regime was estimated to be $0.23$, a value above the threshold for bubble separation without coalescing. The deviation of the bubbles from their spherical shape during the bouncing motion is negligible, which is explained by small Bond numbers ($\sim0.003$). 

\begin{center}\
\begin{figure}[h!]
\includegraphics[width=1.\textwidth]{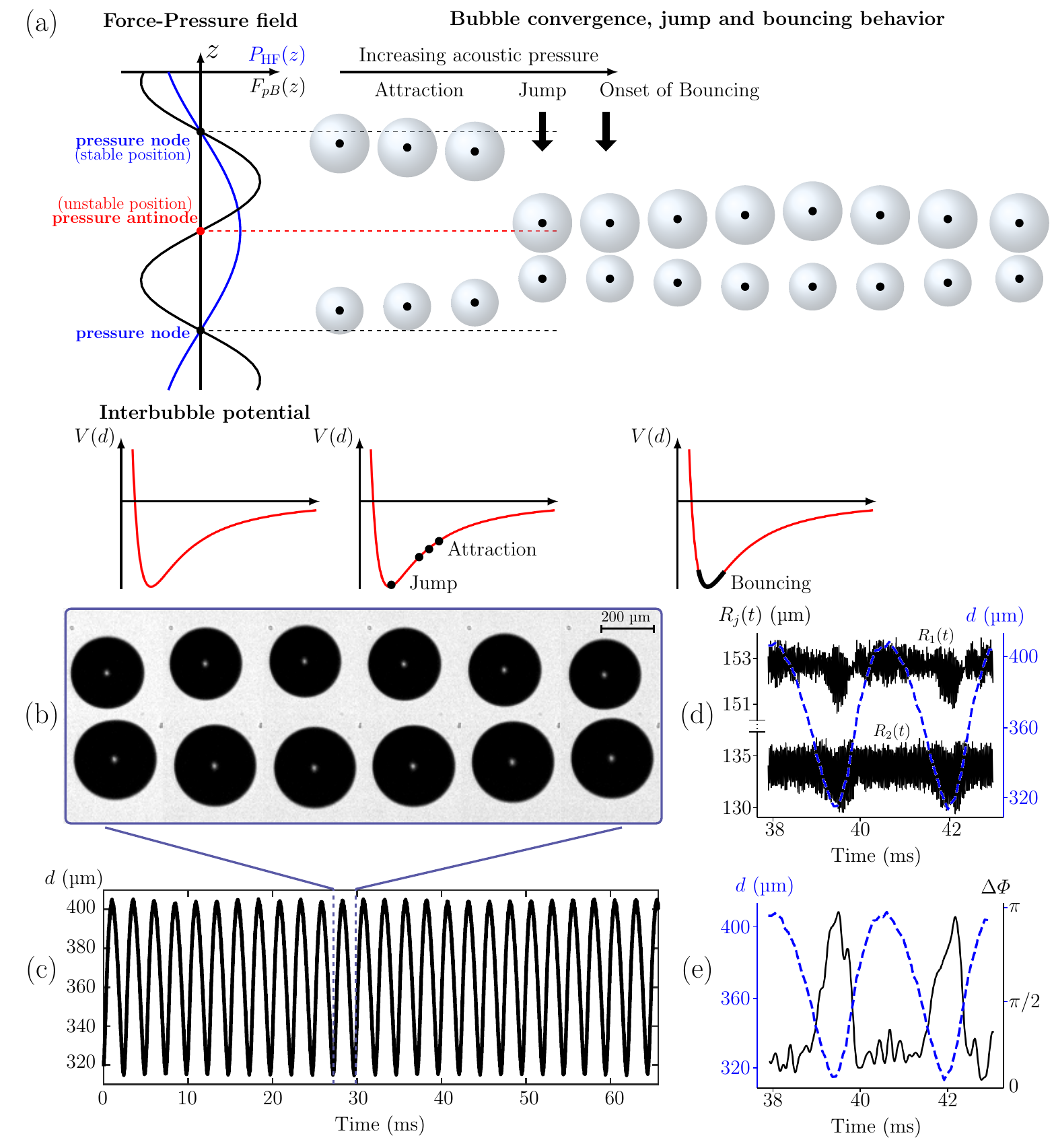}
\caption{\label{fig:fig2}{Observation of a bouncing bubble pair. (a) Illustration of the motion of two approaching bubbles, resulting from their mutual interaction when the applied acoustic pressure is increased. The sign reversal of the interaction radiation force results in the jump phenomenon with the bubbles settling at a very short distance. A further increase in the LF pressure leads to the triggering of the bouncing motion of the bubble pair. The location of the bubble pair in the pressure field is shown as well as its location on the plot of the interbubble potential $V(r)$. (b) The pair of bubbles with radii $R_{10}=$\SI{153}{\micro\meter} and $R_{20}=$\SI{135}{\micro\meter} bounces at the frequency $f_{\textrm{bounce}}=$\SI{403}{\hertz} which is much smaller than the driving frequency. (c) The bouncing is stable during the observation time and the interbubble distance ranges from \SI{315}{\micro\meter} to \SI{405}{\micro\meter}. (d) The bubble oscillations remain spherical during the bouncing. The spherical oscillation (shown by solid black line and denoted by $R_j (t)$, where $R_j (t)$ is the time-varying radius of the $j$th bubble, $j=1,2$) increases significantly when the interbubble distance (blue dash line) is minimal. (e) While the bubbles oscillate in phase at large separation distance (the phase shift $\Delta \Phi \sim 0$, solid black line), they shift to out-of-phase oscillations when the bubbles are at the minimal interbubble distance (blue dash line). Multimedia available online.
}}
\end{figure}
\end{center}\

Because the radii of both bubbles are larger than the LF resonance radius, the bubbles are supposed to oscillate in phase with each other. However, at the minimal interbubble distance, their pulsations increase significantly (Fig.~\ref{fig:fig2}d) and the bubbles oscillate out of phase (the phase shift between the oscillations of the bubbles reaches $\pi$; Fig.~\ref{fig:fig2}e). Note that the bubble pulsations, although of small amplitude, are not harmonic. Similar bouncing phenomena have been observed for bubbles pairs with different equilibrium radii (see Table~\ref{table1}).

\textit{Modeling of the bouncing motion}~--- 
The common description of the dynamics of two interacting pulsating bubbles relies on the assumption that the bubble radii are small compared to the interbubble distance. Since in the present experiments the thickness of the interbubble fluid layer goes down to \SI{30}{\micro\meter} as the bubbles reach the quasi-contact (Fig.~\ref{fig:fig2}b), this assumption does not hold. This suggests that the bubble interaction should be considered at arbitrary (up to very short) separation distance. In the present case, the bubbles are subjected to a bi-frequency acoustic field that includes the HF standing wave field for trapping the bubbles and the LF field that drives the bubble oscillations. Equations for the translation motion of two interacting bubbles undergoing radial pulsations~\cite{Oguz,Barbat,Harkin,Doinikov19,Doinikov20} can be re-written in this case as
\begin{equation}\label{eq:eq1}
\small \rho (R_j \ddot{R_j} + \frac{3}{2} \dot{R_j}^2 - \frac{1}{4} \dot{z_j}^2) - P_j = \sum_{n=1}^{\infty} \frac{p_n^{(j)} ( R_{1,2} , \dot{R_{1,2}} , \ddot{R_{1,2}} , \dot{z_{1,2}} , \ddot{z_{1,2}} )}{d^n},
\end{equation}
\begin{eqnarray}
\frac{d}{dt}(M_j \dot{z_j}) & = & F_{Dj} + F_{bj} + F_{pBj} \nonumber\\
& + & \sum_{n=2}^{\infty} \frac{f_n^{(j)} ( R_{1,2} , \dot{R_{1,2}} , \ddot{R_{1,2}} , \dot{z_{1,2}} , \ddot{z_{1,2}} )}{d^n}, \label{eq:eq2}
\end{eqnarray}
where $j=1,2$, $\rho$ is the liquid density, the overdot denotes the time derivative, $P_j$ is the pressure inside the $j$th bubble, given by~\citep{vanderMeer} $P_j = (P_0 + 2\sigma/R_{j0})(R_{j0}/R_j)^{3\gamma}(1-3\gamma\dot{R_j}/c) - 4\eta\dot{R_j}/R_j - 2\sigma/R_j - P_0 - P_{\textrm{ac}}(t)$, $P_0$ is the liquid hydrostatic pressure, $\sigma$ is the surface tension, $\gamma$ is the ratio of specific heats of the gas, $c$ is the sound speed in the liquid, $\eta$ is the liquid dynamic viscosity, $P_{\textrm{ac}}(t) = -P_a^\textrm{LF}\sin(\omega_{\textrm{LF}t})$ is the LF acoustic pressure, $P_a^\textrm{LF}$ and $\omega_{\textrm{LF}}$ are the pressure amplitude and the angular frequency, respectively, $M_j=(2/3) \pi \rho R_j^3$ is the virtual (added) mass of the $j$th bubble~\cite{Crowe}, $F_{Dj}=-C_{Dj} \dot{z_j}$ is the viscous drag force on the $j$th bubble, $F_{bj}=(4/3)\pi \rho g R_j^3$ is the buoyancy, $g$ is the gravity, $F_{pBj}$ is the primary Bjerknes force exerted on the $j$th bubble by the HF standing wave, given by~\citep{Doinikov2003}
\begin{equation}\label{eq:eqBjerk1}
F_{pBj}(z_j) = \frac{\pi (P_a^\textrm{HF})^2 R_{j0} (\frac{\omega_j^2}{\omega_{\textrm{HF}^2}}-1)\sin (2 k_{\textrm{HF}}z_j)}{\rho\,c\,\omega_{\textrm{HF}} \left[ (\cfrac{\omega_j^2}{\omega_{\textrm{HF}^2}}-1)^2 +(k_{\textrm{HF}} R_{j0})^2 \right]},
\end{equation}
where $P_a^\textrm{HF}$, $\omega_{\textrm{HF}}$ and $k_{\textrm{HF}} = \frac{\omega_{\textrm{HF}}}{c}$ are the pressure amplitude, the angular frequency, and the wavenumber of the HF wave, respectively, and $\omega_j=\sqrt{ 3\gamma P_0 / \rho R_{j0}^2 + 2 (3\gamma - 1)\sigma / \rho R_{j0}^3 }$ is the angular resonance frequency of the $j$th bubble, and the functions $p_n^{(j)},f_n^{(j)}$ are related, respectively, to the additional pressure generated in the liquid due to the interaction of the bubbles and to the radiation interaction force between the bubbles. 

In order to consider the dependence of the drag coefficient $C_{Dj}$ on the Reynolds number $\textrm{Re}= 2 R_j \lvert \dot{z_j} \rvert \rho / \eta$, we use the following approximation. For a rigid sphere, the dependence of $C_{Dj}$ on $\textrm{Re}$ is given by an experimental curve called the standard drag curve (SDC)~\citep{Clift}. Experiments show that $C_{Dj}$ for bubbles in an untreated (tap) water follows the SDC for $\textrm{Re}<10$, while for $\textrm{Re}>10$, the curve for bubbles goes above the SDC~\citep{Levich}, depending on the extent of water contamination~\citep{Gaudin}. To approximate the SDC, we used an equation introduced by Brauer and Mewes~\citep{Clift,Brauer}: $C_{Dj} = \pi \eta R_j (6 + C_1 \sqrt{\textrm{Re}} + C_2 \textrm{Re})$. For $C_1=1$ and $C_2=0.1$, this equation approximates the SDC up to $\textrm{Re}<3\times10^5$. Increasing $C_1$ and $C_2$, one can approximate experimental data for bubbles in an untreated water. Since there is no universal viscous law for bubbles in a wide range of $\textrm{Re}$, it is appropriate to consider that $C_1$ and $C_2$ are fitting parameters. In an exhaustive description, the bubbles also experience one more force, called the Basset (history) force~\citep{Crowe}. However, this force is commonly neglected unless a body is accelerated at a high rate~\citep{Clift,Jonson}, and can thus be disregarded here. Note also that the contribution of the Basset force can be approximated by varying constants in the viscous law. Finally, it should be mentioned that Eqs.~(\ref{eq:eq1}) and (\ref{eq:eq2}), by virtue of their derivation, include hydrodynamic interactions between bubbles~\citep{Oguz}.

The sum in the right-hand side of Eq.~(\ref{eq:eq1}) describes an additional pressure generated in the liquid due to the interaction of the bubbles. It provides the acoustic coupling between the bubble pulsations. The terms of higher order than $d^{-1}$ arise due to multiple scattering of sound between the bubbles and because the scattered waves are produced not only by the radial oscillations but also by the translational components. If $d \gg R_{j0}$, only the first order term is relevant, which corresponds to the single scattering produced by the radial oscillations. The sum in the right-hand side of Eq.~(\ref{eq:eq2}) is the radiation interaction force between the bubbles. The first term of this sum, being averaged over time, recovers the classical secondary Bjerknes force~\citep{Bjerknes}, which describes the interaction of the bubbles for large inter-bubble distances $d \gg R_{j0}$. The terms of higher order in $d^{-1}$ result from sound scattering of higher order and the translational motion. The functions $p_n^{(j)}$ and $f_n^{(j)}$ can be calculated for any n by a procedure developed in Ref.~\citep{Doinikov20}.

Our numerical simulations show that the bouncing bubbles are so close that terms up to $d^{-15}$ should be kept in Eqs.~(\ref{eq:eq1},\ref{eq:eq2}) to accurately describe the multiple sound scattering between the bubbles. In this case, however, the equations of motion are very cumbersome and the modeling is time-consuming. Therefore, we use a simplified model, which nevertheless allows one to obtain reliable estimations for the main parameters of the bouncing motion (the frequency and the amplitude of the translational motion). Our numerical analysis reveals that acceptable modeling accuracy is achieved if Eqs.~(\ref{eq:eq1}) and (\ref{eq:eq2}) are taken up to terms of the order $d^{-2}$ and $d^{-3}$, respectively. Doing so and using the results of Ref.~\citep{Doinikov20}, the following simplified equations are obtained:

\begin{eqnarray}
R_j \ddot{R_j} + \frac{3}{2}\dot{R_j}^2 - \frac{1}{4}\dot{z_j}^2 & - & \frac{P_j}{\rho} = \frac{\alpha_1^{(j)}}{d} + \frac{\alpha_2^{(j)}}{2d^2}, \label{eq:Eq3} \\
\ddot{z_j} + \beta_1^{(j)} \dot{z_j} + \frac{\beta_2^{(j)}}{d^3} \dot{z}_{3-j}  & = & \frac{F_{pBj}(z_j) + F_{bj}}{M_j} \nonumber\\
& + & \frac{\beta_3^{(j)}}{d^2} + \frac{\beta_4^{(j)}}{d^3} , \label{eq:Eq4}
\end{eqnarray}
\begin{eqnarray}
\alpha_1^{(j)} & = & - R_{3-j} (2 \dot{R}_{3-j}^2 + R_{3-j} \ddot{R}_{3-j}), \label{eq:EqA1}\\
\alpha_2^{(j)} & = & - (-1)^j  R_{3-j}^2 \left[ \dot{R}_{3-j} (\dot{z_j} + 5 \dot{z}_{3-j}) + R_{3-j} \ddot{z}_{3-j} \right], \label{eq:A2}\\
\beta_1^{(j)} & = & \frac{3 \dot{R_j}}{R_j} + \frac{C_{Dj}}{M_j}, \label{eq:B1}\\
\beta_2^{(j)} & = & - 3 R_{3-j}^3 \left( \frac{\dot{R_j}}{R_j} + \frac{2 \dot{R}_{3-j}}{R_{3-j}} - \frac{C_{D3-j}}{M_{3-j}} \right), \label{eq:B2}\\
\beta_3^{(j)} & = & 3 (-1)^j R_{3-j}^2 (\frac{\dot{R_1}\dot{R_2}}{R_j} + \frac{2 \dot{R}_{3-j}^2}{R_{3-j}} + \ddot{R}_{3-j}), \label{eq:B3} \\
\beta_4^{(j)} & = & 3 R_{3-j}^3 \frac{F_{pB3-j}(z_{3-j}) + F_{b3-j}}{M_{3-j}}. \label{eq:B4}
\end{eqnarray}

\begin{figure}[t!]
\includegraphics[width=\columnwidth]{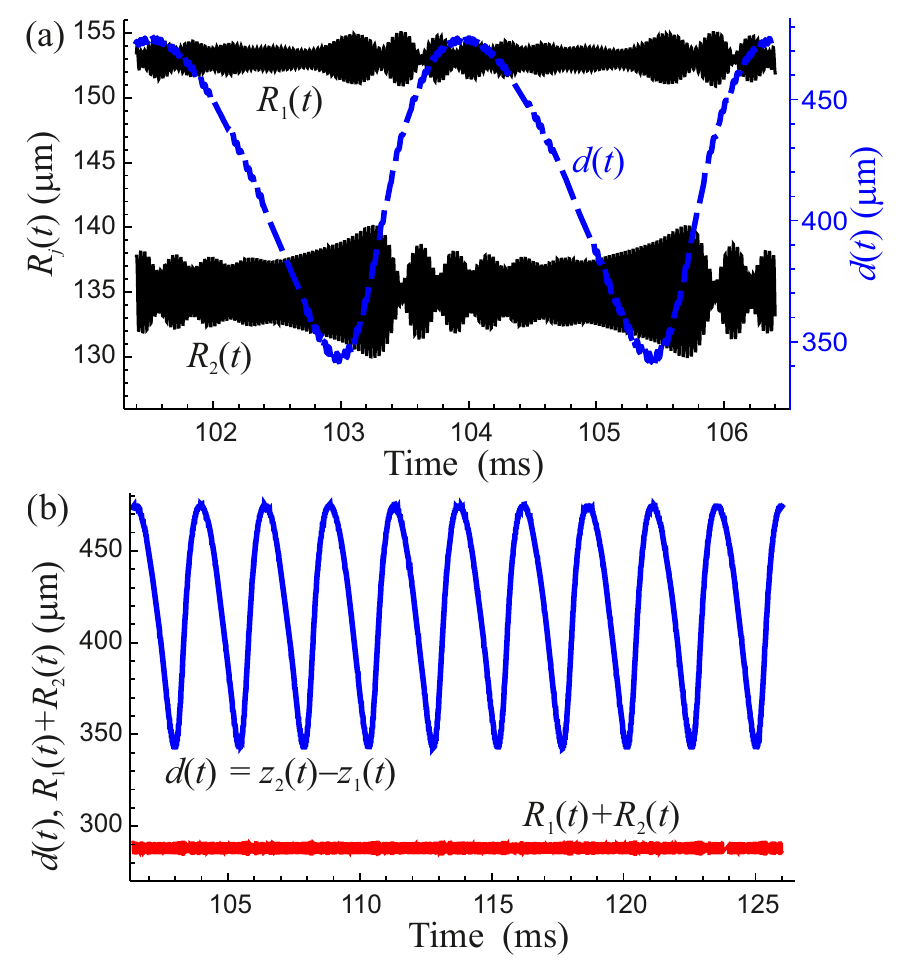}
\caption{\label{fig:fig3}{Theoretical modeling of the dynamics of bouncing bubbles. The numerical simulation of the behavior of two bubbles with initial equilibrium radii $R_{10}=$~\SI{153}{\micro\meter} and $R_{10}=$~\SI{135}{\micro\meter} reveals that the bouncing occurs when the LF pressure exceeds a certain threshold. Here, the bubbles are subjected to a LF acoustic field with $P_a^{\textrm{LF}}=$~\SI{7.3}{\kilo\pascal} and a HF acoustic field with $P_a^{\textrm{HF}}=$~\SI{200}{\kilo\pascal}. The calculations were made by Eqs.~(\ref{eq:Eq3}) and (\ref{eq:Eq4}). (a) The bubble radial oscillations (black solid lines) show a modulational instability, which significantly increases when the interbubble distance (blue dashed line) is minimal. (b) The predicted bouncing behavior is stable over a long time interval.}}
\end{figure}

In the process of the modeling of the driving LF pressure field and the primary Bjerknes force acting on the bubbles, it is necessary to know the LF pressure amplitude $P_a^{\textrm{LF}}$  and the HF pressure amplitude $P_a^{\textrm{HF}}$. The measurement of their exact values is not a trivial experimental task. Previous experiments using the same set-up~\citep{Regnault_PoF} suggest that $P_a^{\textrm{HF}}$ was $\sim$~\SI{200}{\kilo\pascal} and $P_a^{\textrm{LF}} <$~\SI{10}{\kilo\pascal}, so we have set $P_a^{\textrm{HF}}=$~\SI{200}{\kilo\pascal} in our simulations. $P_a^{\textrm{LF}}$, $C_1$, and $C_2$ were allowed to vary in order to reach agreement with the experimental data. For the experimental bouncing shown in Fig.~\ref{fig:fig2}b, our simplified model reproduces the bouncing motion at the frequency $f_{\textrm{bounce}}=$~\SI{403}{\hertz} for $P_a^{\textrm{LF}}=$~\SI{7.3}{\kilo\pascal}, $C_1=1.19$, and $C_2=0.43$. The values of $C_1$ and $C_2$ are reasonable, considering that untreated water was used in the experiments. With these parameters, the main features of the experimental bouncing are recovered (Fig.~\ref{fig:fig3}): the similarity of the general dynamics of the bouncing, the modulated structure of the radial oscillations, and the discernible deviation of the translational oscillation from the harmonic behavior. A good agreement between the experimental and predicted bouncing frequency has been obtained for various bubble pairs with different equilibrium radii (see Table~\ref{table1}).

\begin{table}[h!]
\centering
\caption{Experimental ($f_{\textrm{bounce}}^{\textrm{exp}}$) versus theoretical ($f_{\textrm{bounce}}^{\textrm{th}}$) values of the bouncing frequency for bubble pairs with different equilibrium radii. The bottom row shows the LF pressure at which $f_{\textrm{bounce}}^{\textrm{th}}$ was calculated. The calculations were made by Eqs.~(\ref{eq:Eq3}) and (\ref{eq:Eq4}) for $f_{\textrm{HF}}=$~\SI{1}{\mega\hertz}, $f_{\textrm{LF}}=$~\SI{31.9}{\kilo\hertz}, $P_a^{\textrm{HF}}=$~\SI{200}{\kilo\pascal}, $C_1=1.19$, and $C_2=0.43$. }
\begin{tabular}{lrrrr}
 & Case 1 & Case 2 & Case 3 & Case 4 \\
$R_{10}$ (\SI{}{\micro\meter}) & 153 & 156 & 146 & 145 \\
$R_{20}$ (\SI{}{\micro\meter}) & 135 & 142 & 130 & 129 \\
$f_{\textrm{bounce}}^{\textrm{exp}}$ (\SI{}{\hertz}) & 403 & 340 & 467 & 457 \\
$f_{\textrm{bounce}}^{\textrm{th}}$ (\SI{}{\hertz}) & 403 & 303 & 470 & 447 \\
$P_{a}^{\textrm{LF}}$ (\SI{}{\kilo\pascal}) & 7.3 & 7.4 & 7.4 & 11 \\
\end{tabular}
\label{table1}
\end{table}

\textit{Self-oscillation identification}~--- 
In our experiments, the bouncing was observed only for bubbles driven above resonance ($\omega_{1,2}<\omega_{\textrm{LF}} \Leftrightarrow R_{j0}>R_{\textrm{res}^{\textrm{LF}}}$, where $\omega_j$ is the resonance angular frequency of the $j$th bubble, and $\omega_{\textrm{LF}}$ is the angular frequency of the LF driving acoustic field) and having a sufficiently large difference in radii. No bouncing was observed for almost equal bubbles, which allows one to eliminate the synchronization of two “dancing” bubbles as a possible mechanism for the bouncing motion. Indeed, when a bubble is trapped by a standing acoustic wave, an erratic translational motion, known as "dancing motion", can be triggered above a certain threshold of the driving acoustic pressure~\citep{Eller}. This translational instability occurs at a frequency much smaller than the driving frequency. Therefore, the bouncing motion might be treated as the synchronization of two "dancing" bubbles. This possibility is disregarded by two observations. The first one is the conservation of the spherical shape of two interacting bubbles during the bouncing motion, whereas the "dancing" motion is known to appear at a pressure that corresponds to the triggering of shape oscillations at the bubble interface, which was never the case for the bouncing of spherical bubbles. The second one is the systematic absence of bouncing  when two almost equal bubbles interact. Whereas the synchronization  between oscillators is known to occur only if a small detuning of the natural frequencies of the oscillators is present. It follows that the bouncing motion between two equal-sized bubbles should have existed, which, as said above, is not observed in our experiments. Therefore, the synchronization between two "dancing" bubbles cannot be considered as a potential source of bouncing. 

The large difference between the bouncing frequency and the frequency of the driving external source suggests the mechanism of self-oscillation as a potential candidate. The key feature of the self-oscillatory motion is the concept of negative damping~\citep{Andronov,Landa}, resulting from a delayed action in a dynamical system. In order to evaluate the evolution of damping of the translational motion, we consider Eq.~(\ref{eq:Eq4}), which at $j = 2$ governs the translation of bubble $2$. It contains terms proportional to $\dot{z_1}$ and $\dot{z_2}$ that are responsible for the damping of the translation. The damping term is characterized by the function
\begin{equation}\label{eq:eq5}
\delta = \frac{1}{\langle \dot{z_2} \rangle} \langle \beta_1^{(2)} \dot{z_2} + \frac{\beta_2^{(2)}}{d^3} \dot{z_1} \rangle,
\end{equation}
where $\langle \rangle$ means the averaging over the time interval $2 \pi / \omega_{\textrm{LF}}$. 

\begin{figure}[t!]
\centering
\includegraphics[width=1\linewidth]{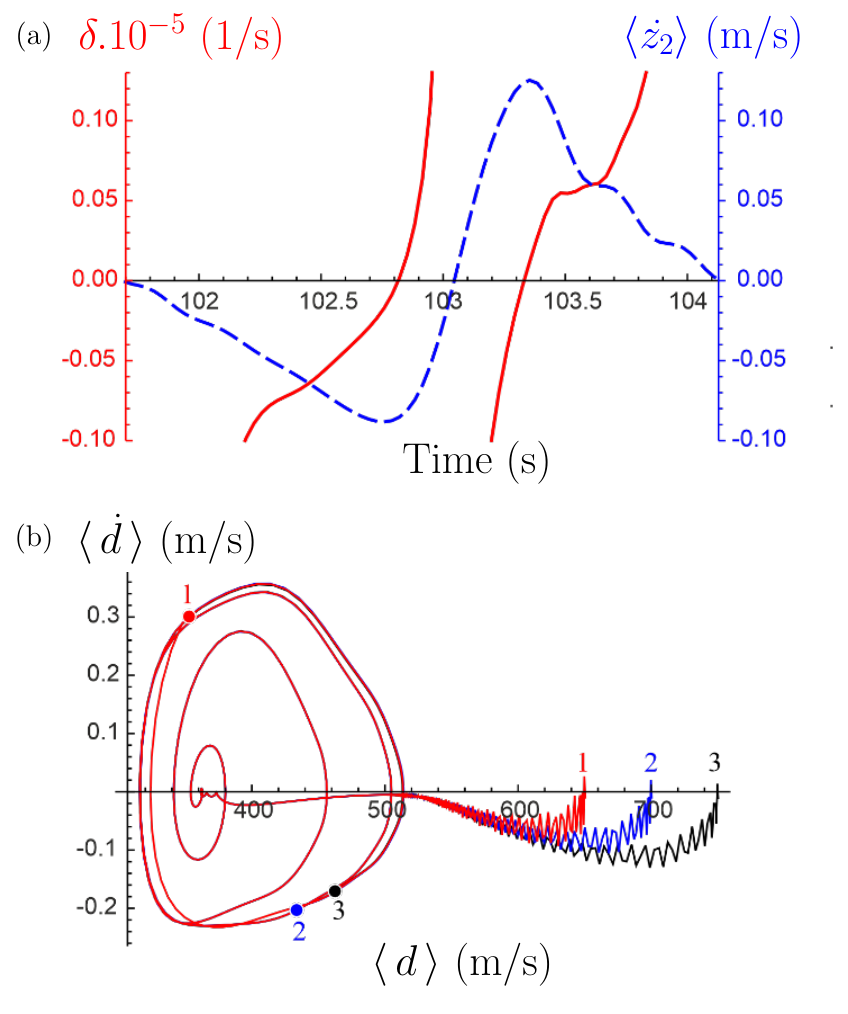}
\caption{Key features of self-oscillatory motion. (a) The transient occurrence of negative damping is numerically observed during one bouncing period. When the magnitude of the translational velocity $\langle \dot{z_2} \rangle$ of bubble 2 (averaged over one acoustic period, blue dashed line) increases, the damping term $\delta$ (red solid line) changes sign. (b) The same limit cycle is reached in the phase plane regardless of the initial conditions (starting points $i=1,2,3$ with zero velocity and different values of $\langle d \rangle$). The calculations were made by Eqs.~(\ref{eq:Eq3}) and ~(\ref{eq:Eq4}) for $f_{\textrm{HF}}=$~\SI{1}{\mega\hertz}, $f_{\textrm{LF}}=$~\SI{31.9}{\kilo\hertz}, $R_{10}=$~\SI{153}{\micro\meter}, $R_{10}=$~\SI{135}{\micro\meter}, $P_a^{\textrm{HF}}=$~\SI{200}{\kilo\pascal}, $P_a^{\textrm{LF}}=$~\SI{7.3}{\kilo\pascal}, $C_1=1.19$, and $C_2=0.43$.}
\label{fig:fig4}
\end{figure}

During one bouncing period (Fig.~\ref{fig:fig4}a), the damping is first negative (which corresponds to amplifying oscillations for a forced harmonic oscillator) and the magnitude of the translational velocity $\langle \dot{z_2} \rangle$ increases. At some moment, the damping becomes positive (decaying oscillations) and the velocity magnitude decreases. Then, the damping becomes negative again, which makes the velocity increase. The damping becomes positive and the velocity decreases again. This process is repeated, providing the stable bouncing. The observed bouncing motion is also characterized by the so-called limit cycle, i.e., a closed trajectory in the phase plane ($d,\dot{d}$) that is identical for different initial conditions (see Fig.~\ref{fig:fig4}b). The initial conditions do not affect the bouncing, which keeps the phase freedom unlike forced and parametric oscillations. The phase freedom is another fundamental feature of self-oscillation~\citep{Pikovsky}.

Physically, the bouncing is explained by the fact that an initial attractive force between the two bubbles reverses to repulsion at a small separation. It is known that two bubbles pulsating in phase attract each other, whereas two bubbles pulsating out of phase repel each other~\citep{Zabolotskaya}. During one bouncing period (Fig.~\ref{fig:fig2}e), the bubbles tend to pulsate in phase at the maximum separation distance and out of phase at the minimum separation distance, which confirms the alternating changeover of the interaction force from attraction to repulsion. The exact moment at which the radiation interaction force reverses in sign corresponds to the equilibrium of the quasi-contacting bubble pair (see Fig~\ref{fig:fig2}a). This location corresponds to the well of the potential energy of the interaction. A further increase in the applied acoustic pressure allows the 2-bubble system, initially trapped in this energy well, to undergo an oscillatory translational motion around the minimum of the potential, so the bouncing motion appears as an analog of the vibrational motion of a diatomic molecule~\citep{Maitland}. It is also worth noting that the bubbles’ out-of-phase oscillations indicate that the antisymmetric mode of the radial pulsations has been triggered during the bouncing motion (see Appendix A). For the particular case $R_{10}=R_{20}$, this mode cannot be excited as two equal-sized bubbles always pulsate in phase if the interbubble distance is small compared to the acoustic wavelength~\citep{Ida}, preventing the existence of bouncing.

It is worth noting that a periodic translational motion of two bubbles at a large separation distance ($14$ to $29$ bubble radii) has been observed by Barbat et al.~\citep{Barbat}. However, our analysis reveals that their modeling is unable to describe our findings, neither the bouncing at very short distances nor the self-oscillation properties. Indeed, when investigating this motion, the authors took into account only the single scattering produced by radial oscillations and terms to $d^{-3}$ in the interaction force. Moreover, they assumed the virtual mass to be constant and neglected $F_{Dj}+F_{bj}+F_{pBj}$. These simplifications lead to a relatively simple model, which predicts a translational oscillation at large bubble distances. This oscillation has nothing to do with the self-oscillation observed in our experiments at short interbubble distances, where the separation at the moment of the bubble convergence is less than $2.5$ radii of the bigger bubble. The model of Barbat et al. is unable to describe the present findings. First, damping is known to be the mandatory element of a self-oscillatory system, whereas the inclusion of the viscous drag into Barbat et al.'s model results in anharmonic or dying oscillations, i.e., their model cannot predict sustained oscillations in the presence of damping. Second, our analysis reveals that the assumption of the single scattering and the constant virtual mass leads to the impossibility of negative damping, which is another key of a self-oscillatory system.

\textit{Discussion}~--- 
The understanding of the dynamics of interacting acoustic bubbles is a challenging physical task that is interesting from a fundamental point of view as well as for practical applications. The bubble-bubble interactions, and, in particular, the mutual interaction force experienced by bubbles oscillating nearby each other (the so-called secondary Bjerknes force), are well known for large inter-bubble distances. Two bubbles oscillating in-phase experience an attractive interaction force while the force is repulsive for out-of-phase oscillations~\citep{Lauterborn}. At a short range, the multiple sound scattering between two close bubbles may lead to the sign reversal of the interaction force, resulting in an attractive-to-repulsive transition in the translational motion of the two bubbles. Ultimately, this sign reversal can lead to the appearance of a new equilibrium location of the two bubbles, which come into near contact~\citep{Regnault_PoF}. This explains the physics of bubble grapes~\citep{Marston}, i.e. stable bubble clusters where the mean separation distance between bubbles is comparable to their sizes, and the self-organization of the structure of cavitation bubble fields~\citep{Parlitz}. Recently, the peculiar properties of the repulsive force acting in bubble aggregates have been used for controlling bubble-propelled micromotors~\citep{Moo}. Here, we show that, in addition to the motion of two interacting bubbles to a quasi-contacting location, they can experience a periodic and stable bouncing motion above a certain threshold of the applied acoustic pressure. This bouncing motion demonstrates all the properties of self-oscillation, including the oscillation at a frequency much smaller than the driving acoustic frequency, the phase freedom and the existence of a limit cycle in the phase plane of the translational motion. To our knowledge, this is the first experimental observation of self-oscillation in the physics of bubbles, which gives a new insight into the understanding of short-range bubble interactions. In addition to the description of the structure formation of cavitation fields, these short-range phenomena can find practical applications in the synchronization of an ensemble of cavitating bubbles. Being a self-oscillator, the bouncing phenomenon possesses a free phase that can be adjusted either by an external driving or matched to the frequency of another bouncing bubble pair in its vicinity through energy transfer between two self-oscillators~\citep{Landa}. Amongst other possibilities, the design of a synchronized bouncing-bubble chain with space and time modulated properties can serve as a microfluidic nonreciprocal system.

The bouncing of bubbles can be used as a fluid micromixer through the generation of acoustic microstreaming, i.e. a slow mean flow induced by the fast oscillations of the bubble interface in a viscous fluid. While a single bubble that pulsates radially in an unbounded fluid does not produce any microstreaming~\citep{Doinikov19b}, the presence of a second bubble and the induced multiple sound scattering leads theoretically to a non-zero microstreaming flow~\citep{Doinikov22}. However, in the case of bubble volume oscillations without translational motion, this flow remains negligible in comparison to that of bubbles experiencing a translational motion, which is a more efficient stream source~\citep{Regnault21}. Using a theoretical modeling of the microstreaming produced by two bubbles experiencing arbitrary interface oscillations, including the volume and translational oscillations at any separation distances~\citep{Doinikov22}, the bouncing bubbles should generate a flow with a maximal velocity of \SI{600}{\micro\meter/\second}, which is ten times higher than the one created by the same bubble pair experiencing only spherical oscillations (see Appendix B).

\textit{Conclusion}~--- 
The investigation of two interacting acoustic bubbles at a short separation distance reveals that above a threshold pressure amplitude, the bubble pair starts a bouncing motion with a bouncing frequency much lower than the driving frequency of the radial oscillations. This motion exhibits all the features of self-oscillation, including negative damping, the phase freedom and the emergence of a limit cycle. Our results highlight the universal nature of the phenomenon of self-oscillation, which, as shown here, can occur even in bubble dynamics, a field far from electronics, where this phenomenon was originally discovered. In a more particular context, our results are relevant for a better understanding and modeling of the dynamics of dense bubble clusters, which are widespread in nature and technology.

\section*{Acknowledgments}
This work was funded by l’Agence Nationale de la Recherche (ANR) , project ANR-22-CE92-0062, and supported by the LabEx CeLyA of the University of Lyon (No. ANR-10-LABX-0060/ANR-11-IDEX-0007).

\appendix*

\section{A. Triggering of the bubbles' antisymmetric oscillations}
\begin{figure}[t!]
\centering
\includegraphics[width=1\linewidth]{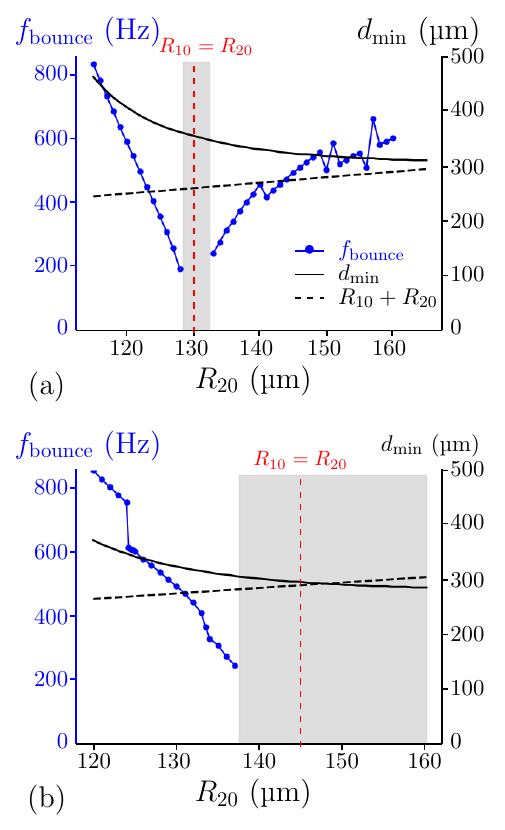}
\caption{Correlation between the appearance of bouncing and the triggering of the bubbles' antisymmetric oscillations. (a) In the case of the fixed $R_{10}=$~\SI{130}{\micro\meter} and varying $R_{20}$, bouncing can occur for a large range of bubble radii as the minimal interbubble distance $d_{\textrm{min}}$, for which the acoustic frequency $\omega$ exceeds that of the antisymmetric mode $\omega_+$, is always larger than the sum of the bubble radii. For $R_{10}\sim R_{20}$, bubbles always oscillate in phase, so that the antisymmetric mode cannot be triggered, preventing the bouncing. (b) In the case of the fixed $R_{10}=$~\SI{145}{\micro\meter} and varying $R_{20}$, the absence of bouncing is observed for $R_{20}>$~\SI{140}{\micro\meter}. At $R_{10}\sim R_{20}$, $d_{\textrm{min}}$ becomes lower than the sum of the bubble radii, preventing the occurrence of the antisymmetric mode as the bubbles cannot coalesce.
}
\label{fig:fig5}
\end{figure}

The system of two spherically-oscillating bubbles possesses the following natural frequencies: 
\begin{equation}\label{eq:eqAppend}
\omega_{\pm}^2 = \cfrac{ \omega_1^2+\omega_2^2 \pm \sqrt{ (\omega_1^2-\omega_2^2)^2 + 4 \omega_1^2 \omega_2^2 R_{10}^2R_{20}^2 / d^2} }{ 2 (1 - R_{10}R_{20}/d^2) },
\end{equation}
where $\omega_-$ ($\omega_+$) is the angular frequency of the symmetric (antisymmetric) mode, where both bubbles oscillate in phase (out-of-phase). For a given bubble pair with known equilibrium radii, Eq.~(\ref{eq:eqAppend}) allows one to define the minimal interbubble distance $d_{\textrm{min}}$ for which the antisymmetric mode is excited, for instance, by considering when the acoustic frequency $\omega$ is close to or exceeds $\omega_+$ (here we consider $\omega=\omega_+$ for predicting $d_{\textrm{min}}$). Numerical investigations of the bouncing behavior, when $R_{10}$ is kept constant while $R_{20}$ varies, confirm that the bouncing is always absent when the bubbles are of almost equal size (see Fig.~\ref{fig:fig5}). In this case, $\omega_+$ coincides with the antiresonance frequency (at which the response of an oscillator sharply falls) of the 2-bubble system, so the antiresonance and the $\omega_+$ modes cancel each other. Also, the absence of bouncing for particular bubble pairs is correlated to the impossibility of triggering the antisymmetric mode, for instance, in the case of the fixed $R_{10}=$~\SI{145}{\micro\meter} and varying $R_{20}$, for the values $R_{20}>$~\SI{140}{\micro\meter}. In that range of $R_{20}$, $d_{\textrm{min}}$ becomes lower than the sum of the bubble radii, preventing the occurrence of antisymmetric modes as the bubbles cannot coalesce. It is worth noting that the triggering of the antisymmetric oscillations of the bubble pair underlies a super-resonance phenomenon~\citep{Feuillade}. The super-resonant mode is characterized by a significant decrease of the radial oscillations of the bubble pair, a high quality factor, meaning that the modal damping is drastically reduced. The associated increase in the radial oscillations feeds the nonlinear coupling effects appearing in the damping term $\delta$ and leads to a transient significant decrease in the damping.  

\section{B. Prediction of the microstreaming induced by the bouncing bubbles}
The fast oscillation of an acoustic bubble is the source of a fluid flow that is slow in comparison to the acoustic timescale. The resulting steady circulation of the fluid is called acoustic microstreaming. The physical origin of acoustic microstreaming is either the interaction between at least two modes of the bubble oscillations (which include the spherical mode, the bubble center translational oscillation or any nonspherical oscillation occurring along the bubble contour) or the interaction of a mode with itself. Because of the time-averaging process occurring in the derivation of the mean flow, nonzero contributions to acoustic microstreaming can only come either from pairs of modes that oscillate at the same frequency, or from the interaction of a mode with itself~\citep{Doinikov19b}. For two bouncing bubbles, the main contributions of the spherical and translational motions occur at two different timescales, so the interfaces of the two bubbles can be written:
\begin{equation}\label{eq:eq7}
R_1(t) = R_{10} + a_{01} e^{\imath \omega t} + a_{11} e^{\imath \omega_b t},
\end{equation}
\begin{equation}\label{eq:eq8}
R_2(t) = R_{20} + a_{02} e^{\imath \omega t + \phi_0} + a_{12} e^{\imath \omega_b t + \phi_b},
\end{equation}
where $a_{0i}$ and $a_{1i}$ are, respectively, the amplitude of the spherical and translational motion of the $i$th  bubble, $\omega$ and $\omega_b$ are, respectively, the acoustic and bouncing angular frequencies, and $\phi_0,\phi_b$ are, respectively, the phase delay between the bubble spherical oscillations and between their translational motion occurring at the bouncing frequency. Since only oscillations occurring at the same frequency can produce microstreaming, contributions to the mean flow come only from the interaction of the spherical oscillations of the bubbles at the frequency $\omega$ and from the interaction of the translation oscillations of the bubbles at the frequency $\omega_b$. Using the theoretical modelling accounting for the interaction of two bubbles oscillating at any separation distances~\citep{Doinikov22}, one can thus predict the order of magnitude of the induced mean flow for both interactions. Neglecting the slowly-varying envelope of the spherical oscillations (Fig.~\ref{fig:fig2}d), the following values have been used for predicting the mean flow induced by spherical oscillations alone: $a_{01} = a_{02}=$~\SI{1}{\micro\meter}, $\omega=2\pi \times 31.9 10^3$\SI{}{\radian/\second}, $\phi_0=0$, and the interbubble distance $d=$~\SI{360}{\micro\meter}. The associated Lagrangian streaming velocity reaches a maximum absolute velocity of \SI{44}{\micro\meter/\second}. Neglecting the fast-varying components of the translational oscillations, the following values have been used for predicting the mean flow induced by the translational motion alone: $a_{11} = a_{12} =$~\SI{15}{\micro\meter}, $\omega_b = 2\pi \times 403$ \SI{}{\radian/\second}, and $d=$~\SI{360}{\micro\meter}. The bouncing motion is also characterized by $\phi_b=\pi$. The amplitude of the translational motion has been reduced in comparison to the observed one (around ~\SI{40}{\micro\meter} in Fig.~\ref{fig:fig2}) in order to follow the assumption of small translation displacement in comparison to the mean equilibrium radius. The associated Lagrangian streaming velocity reaches a maximum absolute velocity of about \SI{600}{\micro\meter/\second}.

\bibliographystyle{elsarticle-num}
\bibliography{bib_file}

\end{document}